\documentclass[12pt]{article}
\usepackage{amsfonts}

\usepackage{graphicx}
\usepackage{amsmath}
\usepackage{float}

\textwidth=6.5in \hoffset=-.75in \voffset=-1in \numberwithin{equation}{section}
\numberwithin{figure}{section}

\input{tcilatex}

\begin{document}

\begin{titlepage}
\bigskip \begin{flushright}
%hep-th/0606210\\
\end{flushright}
%\maketitle
\vspace{1cm}
\begin{center}
{\Large \bf {Bianchi Type IX M-Branes}}\\
\end{center}
\vspace{2cm}
\begin{center}
 A. M.
Ghezelbash{ \footnote{ EMail: amasoud@sciborg.uwaterloo.ca}}
\\
Department of Physics and Astronomy, University of Waterloo, \\
Waterloo, Ontario N2L 3G1, Canada\\
\vspace{1cm}
%PACS numbers:
%\\
%\vspace{2cm}
%\today\\
\end{center}

\begin{abstract}

We present new M2 and M5 brane solutions in M-theory based on transverse 
self-dual Bianchi type IX space. All the other recently M2 and M5 branes constructed 
on transverse self-dual Taub-NUT, Egughi-Hanson and Atiyah-Hitchin spaces are special cases of 
this solution.  The solution provides a smooth transition from Eguchi-Hanson type I based M branes to corresponding branes based on Eguchi-Hanson type II space.
All the solutions can be reduced down to ten dimensional fully localized intersecting  brane configurations. 
\end{abstract}
\bigskip
\hspace*{1cm} PACS: 11.25.Yb,11.25.Uv,11.30.Pb,11.25.-w
\end{titlepage}\onecolumn

\bigskip

\section{Introduction}

Fundamental M-theory in the low-energy limit is generally believed to be
effectively described by $D=11$ supergravity \cite{gr1,gr2,gr3}. This
suggests that brane solutions in the latter theory furnish classical soliton
states of M-theory, motivating considerable interest in this subject. There
is particular interest in supersymmetric $p$-brane solutions that saturate
the Bogomol'nyi-Prasad-Sommerfield (BPS) bound upon reduction to 10
dimensions. Some supersymmetric solutions of two or three orthogonally
intersecting 2-branes and 5-branes in $D=11$\ supergravity were obtained
some years ago \cite{Tsey}, and more such solutions have since been found %
\cite{other}.

Recently interesting new supergravity solutions for localized D2/D6, D2/D4,
NS5/D6 and NS5/D5 intersecting brane systems were obtained \cite%
{hashi,CGMM2,ATM2}. By lifting a D6 (D5 or D4)-brane to four-dimensional
Taub-NUT/Bolt, Eguchi-Hanson and Atiyah-Hitchin geometries embedded in
M-theory, these solutions were constructed by placing M2- and M5-branes in
the Taub-NUT/Bolt, Eguchi-Hanson and Atiyah-Hitchin background geometries.
The special feature of these constructions\ is that the solution is not
restricted to be in the near core region of the D6 (D5 or D4)-brane.\ \ 

Taub-NUT, Eguchi-Hanson (the latter space will be refered to as
Eguchi-Hanson type II in this paper) and Atiyah-Hitchin spaces are each
special cases of the Bianchi type IX space. \ The Bianchi type IX spaces
were used recently for construction cohomogeneity two metrics of G$_{2}$
holonomy which are foliated by twisteor spaces \cite{CV}. The twistor spaces
are two-sphere bundles over Bianchi type IX Einstein metrics with self-dual
Weyl tensor.

Since the building blocks of M-theory are M2- and M5-branes, it is natural
to investigate the possibility of placing M2- and M5-branes in the Bianchi
type IX background space. This is the subject of the present paper, in which
we consider the embedding of\ \ Bianchi type IX geometry in M-theory with an
M2- or M5-brane.\ For all of the different solutions we obtain, 1/4 of the
supersymmetry is preserved as a result of the self-duality of the Bianchi
type IX metric. We then compactify these solutions on a circle, obtaining
the different fields of type IIA string theory. Explicit calculation shows
that in all cases the metric is asymptotically (locally) flat, though for
some of the compactified solutions the type IIA dilaton field diverges at
infinity.

The outline of this paper is as follows. In section \ref{sec:review}, we
discuss briefly the field equations of supergravity, the M2- and M5-brane
metrics and the Killing spinor equations. In section \ref{sec:M2}, we
present the different M2-brane solutions that preserve 1/4 of the
supersymmetry. We find type IIA D2$\perp $D6(2) intersecting brane solutions
upon dimensional reduction. \ In section \ref{sec:sc} the alternative
M2-brane solutions are presented. These solutions are obtained by
continuation of the real separation constant into a pure imaginary
separation constant. In section \ref{sec:M5}, we present different M5
solutions that preserve 1/4 of the supersymmetry. In section \ref{sec:declim}%
, we consider the decoupling limit of these solutions, especially M2 brane
solutions.

\section{M2- and M5- Branes and Kaluza-Klein Reduction}

\label{sec:review}

The equations of motion for eleven dimensional supergravity when we have
maximal symmetry (i.e. for which the expectation values of the fermion
fields is zero), are \cite{DuffKK}%
\begin{eqnarray}
R_{mn}-\frac{1}{2}g_{mn}R &=&\frac{1}{3}\left[ F_{mpqr}F_{n}^{\phantom{n}%
pqr}-\frac{1}{8}g_{mn}F_{pqrs}F^{pqrs}\right]  \label{GminGG} \\
\nabla _{m}F^{mnpq} &=&-\frac{1}{576}\varepsilon ^{m_{1}\ldots
m_{8}npq}F_{m_{1}\ldots m_{4}}F_{m_{5}\ldots m_{8}}  \label{dF}
\end{eqnarray}%
where the indices $m,n,\ldots $ are 11-dimensional world space indices. For
an M2-brane, we use the metric and four-form field strength 
\begin{equation}
ds_{11}^{2}=H(y,r)^{-2/3}\left( -dt^{2}+dx_{1}^{2}+dx_{2}^{2}\right)
+H(y,r)^{1/3}\left( d\mathfrak{s}_{4}^{2}(y)+ds_{4}^{2}(r)\right)
\label{ds11genM2}
\end{equation}%
and non-vanishing four-form field components%
\begin{equation}
F_{tx_{1}x_{2}y}=-\frac{1}{2H^{2}}\frac{\partial H}{\partial y}~\
~,~~~F_{tx_{1}x_{2}r}=-\frac{1}{2H^{2}}\frac{\partial H}{\partial r}.
\end{equation}%
and for an M5-brane, the metric and four-form field strength are

\begin{eqnarray}
ds^{2} &=&H(y,r)^{-1/3}\left( -dt^{2}+dx_{1}^{2}+\ldots +dx_{5}^{2}\right)
+H(y,r)^{2/3}\left( dy^{2}+ds_{4}^{2}(r)\right) ~~~  \label{ds11general} \\
F_{m_{1}\ldots m_{4}} &=&\frac{\alpha }{2}\epsilon _{m_{1}\ldots
m_{5}}\partial ^{m_{5}}H~\ ~,~~~\alpha =\pm 1  \label{Fgeneral}
\end{eqnarray}%
where $d\mathfrak{s}_{4}^{2}(y)$ and $ds_{4}^{2}(r)$ are two
four-dimensional (Euclideanized) metrics, depending on the non-compact
coordinates $y$ and $r$, respectively and the quantity $\alpha =\pm 1,$\
which corresponds to an M5 brane and an anti-M5 brane respectively. The
general solution, where the transverse coordinates are given by a flat
metric, admits a solution with 16 Killing spinors \cite{gauntlett}.

The 11D metric and four-form field strength can be easily reduced down to
ten dimensions using the following equations%
\begin{eqnarray}
g_{mn} &=&\left[ 
\begin{array}{cc}
e^{-2\Phi /3}\left( g_{\alpha \beta }+e^{2\Phi }C_{\alpha }C_{\beta }\right)
& \nu e^{4\Phi /3}C_{\alpha } \\ 
\nu e^{4\Phi /3}C_{\beta } & \nu ^{2}e^{4\Phi /3}%
\end{array}%
\right]  \label{FKKreduced1} \\
F_{(4)} &=&\mathcal{F}_{(4)}+\mathcal{H}_{(3)}\wedge dx_{10}
\label{FKKreduced2}
\end{eqnarray}

Here $\nu $ is the winding number (the number of times the M5 brane wraps
around the compactified dimensions) and $x_{10}$ is the eleventh dimension,
on which we compactify. We use hats in the above to differentiate the
eleven-dimensional fields from the ten-dimensional ones that arise from
compactification. $\mathcal{F}_{(4)}$ and $\mathcal{H}_{(3)}$ are the
Ramond-Ramond (RR) four-form and the Neveu-Schwarz-Neveu-Schwarz (NSNS)
three-form field strengths corresponding to $A_{\alpha \beta \gamma }$ and $%
B_{\alpha \beta }$.

Supersymmetric solutions to the equations of motion (\ref{GminGG}) and (\ref%
{dF}) can be constructed by looking for bosonic backgrounds that admit
Killing spinors {\cite{MS, SMITH}}. For all bosonic backgrounds the
supersymmetry variation of the gravitino field must vanish which yields,

\begin{equation}
\partial _{m}\epsilon +\frac{1}{4}\omega _{abm}\Gamma ^{ab}\epsilon +\frac{1%
}{144}\Gamma _{m}^{\phantom{m}npqr}F_{npqr}\epsilon -\frac{1}{18}\Gamma
^{pqr}F_{mpqr}\epsilon =0  \label{killingspinoreq}
\end{equation}%
where $\epsilon$ is 32-component Majorana Killing spinor. The number of
non-trivial solutions to the Killing spinor equation (\ref{killingspinoreq})
determines the amount of supersymmetry of the solution {\cite{J}}.

In equation (\ref{killingspinoreq}) the $\omega $'s are the spin connection
coefficients and $\Gamma ^{a_{1}\ldots a_{n}}=\Gamma ^{\lbrack a_{1}}\ldots
\Gamma ^{a_{n}]}$. The indices $a,b,...$ are 11 dimensional tangent space
indices and the $\Gamma ^{a}$ matrices are the eleven dimensional
equivalents of the four dimensional Dirac gamma matrices, and must satisfy
the Clifford algebra 
\begin{equation}
\left\{ \Gamma ^{a},\Gamma ^{b}\right\} =-2\eta ^{ab}  \label{cliffalg}
\end{equation}

In ten dimensional type IIA string theory, we can have D-branes or
NS-branes. D$p$-branes can carry either electric or magnetic charge with
respect to the RR fields; the metric takes the form \cite{gauntlett}%
\begin{equation}
ds_{10}^{2}=f^{-1/2}\left( -dt^{2}+dx_{1}^{2}+\ldots +dx_{p}^{2}\right)
+f^{1/2}\left( dx_{p+1}^{2}+\ldots +dx_{9}^{2}\right)  \label{gDpbrane}
\end{equation}%
where the harmonic function\thinspace\ $f$ generally depends on the
transverse coordinates.

An NS5-brane carries a magnetic two-form charge; the corresponding metric
has the form 
\begin{equation}
ds_{10}^{2}=-dt^{2}+dx_{1}^{2}+\ldots +dx_{5}^{2}+f\left( dx_{6}^{2}+\ldots
+dx_{9}^{2}\right)  \label{gNS5brane}
\end{equation}%
In what follows we will obtain a mixture of D-branes and NS-branes.

\section{Embedding triaxial Bianchi type IX space in an M2-brane metric}

\label{sec:M2}

The eleven dimensional M2-brane with an embedded transverse triaxial Bianchi
type IX space is given by the following metric%
\begin{equation}
ds_{11}^{2}=H(y,r)^{-2/3}\left( -dt^{2}+dx_{1}^{2}+dx_{2}^{2}\right)
+H(y,r)^{1/3}\left( dy^{2}+y^{2}d\Omega _{3}^{2}+ds_{Bianchi\text{ }%
IX}^{2}\right)  \label{ds11m2}
\end{equation}%
and non-vanishing four-form field components%
\begin{equation}
F_{tx_{1}x_{2}y}=-\frac{1}{2H^{2}}\frac{\partial H}{\partial y}~\
~,~~~F_{tx_{1}x_{2}r}=-\frac{1}{2H^{2}}\frac{\partial H}{\partial r}.
\label{Fm2}
\end{equation}%
The triaxial Bianchi type IX metric $ds_{Bianchi\text{ }IX}^{2}$ is locally
given by the following metric with an $SU(2)$ or $SO(3)$ isometry group \cite%
{GM} 
\begin{equation}
ds_{Bianchi\text{ }IX}^{2}=e^{2\{A(\zeta )+B(\zeta )+C(\zeta )\}}d\zeta
^{2}+e^{2A(\zeta )}\sigma _{1}^{2}+e^{2B(\zeta )}\sigma _{2}^{2}+e^{2C(\zeta
)}\sigma _{3}^{2}  \label{BIXG}
\end{equation}%
with 
\begin{equation}
\begin{array}{c}
\sigma _{1}=d\psi +\cos \theta d\phi \\ 
\sigma _{2}=-\sin \psi d\theta +\cos \psi \sin \theta d\phi \\ 
\sigma _{3}=\cos \psi d\theta +\sin \psi \sin \theta d\phi%
\end{array}
\label{mcFORMS}
\end{equation}%
where $\sigma _{i\text{ }}$are Maurer-Cartan one-forms with the property 
\begin{equation}
d\sigma _{i}=\frac{1}{2}\varepsilon _{ijk}\sigma _{j}\wedge \sigma _{k}.
\label{dsigma}
\end{equation}%
We note that the metric on the $\mathbb{R}^{4}$ (with a radial coordinate $R$
and Euler angles ($\theta ,\phi ,\psi $) on an $S^{3}$) could be written in
terms of Maurer-Cartan one-forms via%
\begin{equation}
ds^{2}=dR^{2}+\frac{R^{2}}{4}(\sigma _{1}^{2}+\sigma _{2}^{2}+\sigma
_{3}^{2}).  \label{s3METRIC}
\end{equation}%
with $\sigma _{1}^{2}+\sigma _{2}^{2}$\ is the standard metric of $S^{2}$\
with unit radius; $4(\sigma _{1}^{2}+\sigma _{2}^{2}+\sigma _{3}^{2})$\
gives the same for $S^{3}.$ The metric (\ref{BIXG}) satisfies Einstein's
equations provided%
\begin{equation}
\begin{array}{c}
\frac{d^{2}A}{d\zeta ^{2}}=\frac{1}{2}\{e^{4A}-(e^{2B}-e^{2C})^{2}\} \\ 
\frac{d^{2}B}{d\zeta ^{2}}=\frac{1}{2}\{e^{4B}-(e^{2C}-e^{2A})^{2}\} \\ 
\frac{d^{2}C}{d\zeta ^{2}}=\frac{1}{2}\{e^{4C}-(e^{2A}-e^{2B})^{2}\}.%
\end{array}
\label{Ein}
\end{equation}%
\newline
and%
\begin{equation}
\frac{dA}{d\zeta }\frac{dB}{d\zeta }+\frac{dB}{d\zeta }\frac{dC}{d\zeta }+%
\frac{dC}{d\zeta }\frac{dA}{d\zeta }=\frac{1}{2}%
\{e^{2(A+B)}+e^{2(B+C)}+e^{2(C+A)}\}-\frac{1}{4}\{e^{4A}+e^{4B}+e^{4C}\}
\label{Ein2}
\end{equation}%
\bigskip Moreover self-duality of the curvature implies 
\begin{equation}
\begin{array}{c}
\frac{dA}{d\zeta }=\frac{1}{2}\{e^{2B}+e^{2C}-e^{2A}\}-\alpha _{1}e^{B+C} \\ 
\frac{dB}{d\zeta }=\frac{1}{2}\{e^{2C}+e^{2A}-e^{2B}\}-\alpha _{2}e^{A+C} \\ 
\frac{dC}{d\zeta }=\frac{1}{2}\{e^{2A}+e^{2B}-e^{2C}\}-\alpha _{3}e^{A+B}%
\end{array}
\label{sd}
\end{equation}%
where three constant numbers $\alpha _{1},\alpha _{2}$ and $\alpha _{3}$
must satisfy $\alpha _{1}\alpha _{2}=\alpha _{3},\alpha _{2}\alpha
_{3}=\alpha _{1}$ and $\alpha _{3}\alpha _{1}=\alpha _{2}.$

Choosing $(\alpha _{1},\alpha _{2},\alpha _{3})=(1,1,1),$ yields the
Atiyah-Hitchin metric \cite{Hana} in the form of \ (\ref{BIXG}) with%
\begin{equation}
\begin{array}{c}
A=\frac{1}{2}\ln (\frac{2}{\pi }\frac{\vartheta _{2}\vartheta _{3}^{^{\prime
}}\vartheta _{4}^{^{\prime }}}{\vartheta _{2}^{^{\prime }}\vartheta
_{3}\vartheta _{4}}) \\ 
B=\frac{1}{2}\ln (\frac{2}{\pi }\frac{\vartheta _{2}^{^{\prime }}\vartheta
_{3}\vartheta _{4}^{^{\prime }}}{\vartheta _{2}\vartheta _{3}^{^{\prime
}}\vartheta _{4}}) \\ 
C=\frac{1}{2}\ln (\frac{2}{\pi }\frac{\vartheta _{2}^{^{\prime }}\vartheta
_{3}^{^{\prime }}\vartheta _{4}}{\vartheta _{2}\vartheta _{3}\vartheta
_{4}^{^{\prime }}})%
\end{array}
\label{ABC3}
\end{equation}%
where the $\vartheta $'s are Jacobi Theta functions with complex modulus $%
i\zeta .$ Since Atiyah-Hitchin space embedded into the transverse geometry
has been previously used to construct different M2 and M5 brane solutions %
\cite{ATM2}, we don't consider this case here.

By choosing $(\alpha _{1},\alpha _{2},\alpha _{3})=(0,0,0)$ the differential
equations (\ref{Ein}), (\ref{Ein2}) and (\ref{sd}) can be solved exactly to
give a metric of the form%
\begin{equation}
ds^{2}=\frac{dr^{2}}{\sqrt{F(r)}}+\frac{r^{2}}{4}\sqrt{F(r)}\left( \frac{%
\sigma _{1}^{2}}{1-\frac{a_{1}^{4}}{r^{4}}}+\frac{\sigma _{2}^{2}}{1-\frac{%
a_{2}^{4}}{r^{4}}}+\frac{\sigma _{3}^{2}}{1-\frac{a_{3}^{4}}{r^{4}}}\right)
\label{BIX}
\end{equation}%
where%
\begin{equation}
F(r)=\prod_{i=1}^{3}(1-\frac{a_{i}^{4}}{r^{4}})  \label{FF}
\end{equation}%
and $a_{1},a_{2}$ and $a_{3}$ are constants.

The metric (\ref{ds11m2}) with $ds_{Bianchi\text{ }IX}^{2}$ given by (\ref%
{BIX}) is a solution to the eleven dimensional supergravity equations
provided $H\left( y,r\right) $ is a solution to the differential equation 
\begin{equation}
y\sqrt{A_{1}A_{2}A_{3}}[2r\frac{\partial ^{2}H}{\partial r^{2}}+\{6+r(\frac{1%
}{A_{1}}\frac{dA_{1}}{dr}+\frac{1}{A_{2}}\frac{dA_{2}}{dr}+\frac{1}{A_{3}}%
\frac{dA_{3}}{dr})\}\frac{\partial H}{\partial r}]+\{2yr\frac{\partial ^{2}H%
}{\partial y^{2}}+6r\frac{\partial H}{\partial y}\}=0.  \label{tndiffeq}
\end{equation}%
where $A_{i}=1-\frac{a_{i}^{4}}{r^{4}}.$ This equation is straightforwardly
separable. Substituting 
\begin{equation}
H(y,r)=1+Q_{M2}Y(y)R(r)  \label{Hyrsep}
\end{equation}%
where $Q_{M2}$ is the charge on the M2 brane, we arrive at two differential
equations for $Y(y)$ and $R(r).$ One solution of the differential equation
for $Y(y)$ is 
\begin{equation}
Y(y)=\frac{J_{1}(\frac{c}{\sqrt{2}}y)}{y}  \label{Y1}
\end{equation}%
which has the preferred damped oscillating behavior at infinity. The other
solution is the Bessel function of \ the second kind, which diverges at
infinity. The differential equation for $R(r)$ is

\begin{equation}
2rA_{1}A_{2}A_{3}\frac{d^{2}R_{c}(r)}{dr^{2}}+\{6A_{1}A_{2}A_{3}+r(A_{2}A_{3}%
\frac{dA_{1}}{dr}+A_{3}A_{1}\frac{dA_{2}}{dr}+A_{1}A_{2}\frac{dA_{3}}{dr})\}%
\frac{dR_{c}(r)}{dr}-c^{2}r\sqrt{A_{1}A_{2}A_{3}}R_{c}(r)=0  \label{M2eqR}
\end{equation}%
where $c$ is the separation constant.

\begin{figure}[tbp]
\centering                         
\begin{minipage}[c]{.3\textwidth}
        \centering
        \includegraphics[width=\textwidth]{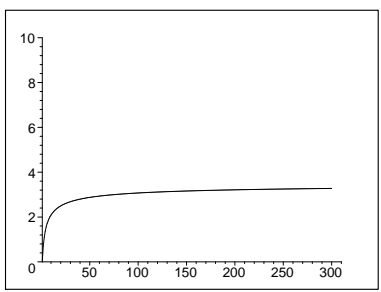}
    \end{minipage}
\caption{ Numerical solution of the radial equation (\ref{M2eqR}) for $%
R_{k}/10^{8}$ as a function of $\frac{1}{r-a_{3}},$ where we set $%
a_{1}=1,a_{2}=2$ and $a_{3}=3$ for simplicity. So for $r\approx a_{3}$ $R$
diverges. Note that the plot is only reliable for large $\frac{1}{r-a_{3}}$.}
\label{M2R123}
\end{figure}

\begin{figure}[tbp]
\centering                         
\begin{minipage}[c]{.3\textwidth}
        \centering
        \includegraphics[width=\textwidth]{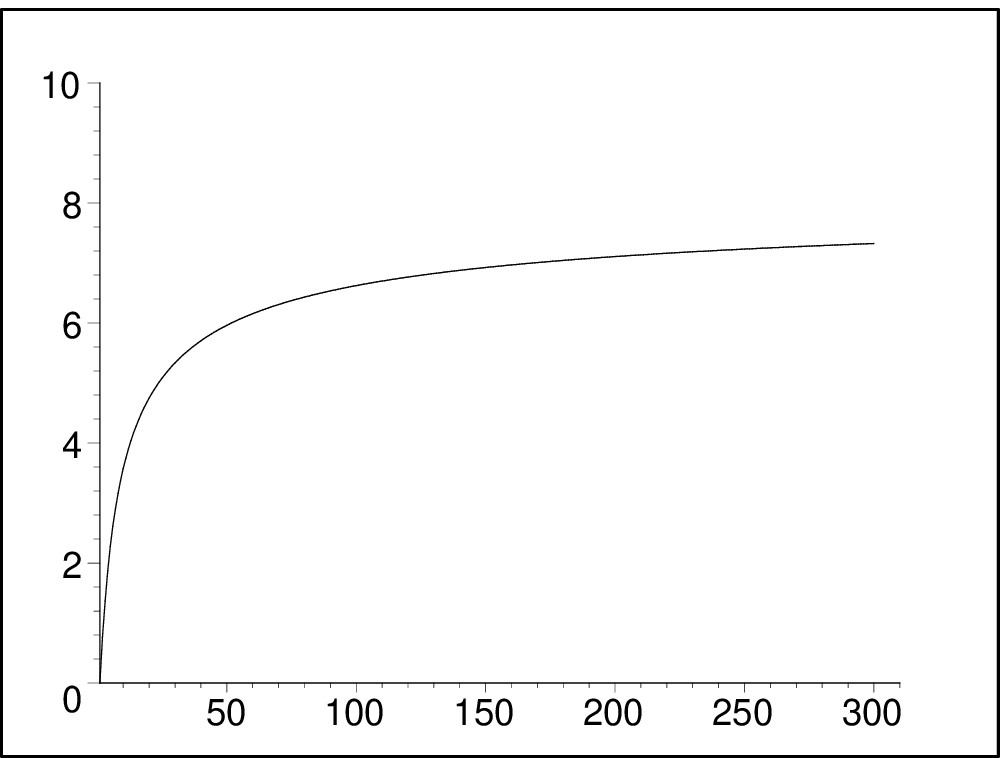}
    \end{minipage}
\caption{ Numerical solution of the radial equation (\ref{M2eqR}) for $%
R_{k}/10^{8}$ as a function of $\frac{1}{r-a_{3}},$ where we set $%
a_{1}=1,a_{2}=2$ and $a_{3}=500$ for simplicity. So for $r\approx a_{3}$ $R$
diverges. Note that the plot is only reliable for large $\frac{1}{r-a_{3}}$.}
\label{M2R12500}
\end{figure}

\begin{figure}[tbp]
\centering                         
\begin{minipage}[c]{.3\textwidth}
        \centering
        \includegraphics[width=\textwidth]{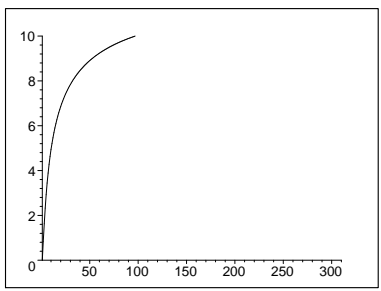}
    \end{minipage}
\caption{ Numerical solution of the radial equation (\ref{M2eqR}) for $%
R_{k}/10^{8}$ as a function of $\frac{1}{r-a_{3}},$ where we set $%
a_{1}=1,a_{2}=2$ and $a_{3}=1000$ for simplicity. So for $r\approx a_{3}$ $R$
diverges. Note that the plot is only reliable for large $\frac{1}{r-a_{3}}$.}
\label{M2R121000}
\end{figure}

Although equation (\ref{M2eqR}) does not have any analytic closed solution,
we can solve it numerically. Several typical numerical solutions of (\ref%
{M2eqR}) are given in figures \ref{M2R123},\ref{M2R12500} and \ref{M2R121000}
for different values of $a_{1},a_{2}$ and $a_{3}$ where we set $a_{1}\leq
a_{2}$ $\leq $ $a_{3}$ and so $r>a_{3}$ in the metric (\ref{BIX}).

We note that by increasing $a_{3}$, the M2 brane metric function for
constant $r$ increases. The most general solution for the metric function
will be a superposition of all possible solutions and takes the form 
\begin{equation}
H_{Bianchi\text{ }IX}(y,r)=1+Q_{M2}\int_{0}^{\infty }\frac{dcc^{4}}{y}J_{1}(%
\frac{c}{\sqrt{2}}y)R_{c}(r)  \label{gensolutionnearreqPi}
\end{equation}%
as the metric function of the M2-brane solution (\ref{ds11m2}) with
transverse Bianchi type IX space, where the measure function is chosen via
dimensional analysis.

An interesting result is obtained by taking the special case%
\begin{equation}
\begin{array}{c}
a_{1}=0 \\ 
a_{2}=2kc \\ 
a_{3}=2c%
\end{array}
\label{special}
\end{equation}%
where we choose $0\leq k\leq 1$ and $c>0.$ For the special value of $k=0,$
where the smaller two $a$'s coincide, we obtain the following metric%
\begin{equation}
ds_{EHI}^{2}=\frac{dr^{2}}{f(r)}+\frac{r^{2}}{4}f(r)\{d\theta ^{2}+\sin
^{2}\theta d\phi ^{2}\}+\frac{r^{2}}{4f(r)}(d\psi +\cos \theta d\phi )^{2}
\label{EHI}
\end{equation}%
which is the Eguchi-Hanson type I metric with $f(r)=(1-\frac{(2c)^{4}}{r^{4}}%
)^{1/2}$. In the other extreme case where $k=1$, the larger two $a$'s
coincide and we obtain the Eguchi-Hanson type II metric%
\begin{equation}
ds_{EHII}^{2}=\frac{dr^{2}}{f^{2}(r)}+\frac{r^{2}}{4}f^{2}(r)\sigma _{1}^{2}+%
\frac{r^{2}}{4}(\sigma _{2}^{2}+\sigma _{3}^{2})  \label{EHII}
\end{equation}%
This met\bigskip ric can be changed to the standard form given in \cite{Ma} 
\begin{eqnarray}
ds_{EH}^{2} &=&\frac{r^{2}}{4g(r)}\left[ d\psi +\cos (\theta )d\phi \right]
^{2}+g(r)dr^{2}+\frac{r^{2}}{4}\left( d\theta ^{2}+\sin ^{2}(\theta )d\phi
^{2}\right) \\
g(r) &=&\left( 1-\frac{a^{4}}{r^{4}}\right) ^{-1}.
\end{eqnarray}%
by making the substitutions $\sigma _{1}\longleftrightarrow \sigma _{3}$ and 
$2c=a$ in (\ref{EHII}). Another form of the Eguchi-Hanson type I metric can
be written as \cite{EH} 
\begin{equation}
ds_{EHI}^{2}=\widetilde{f}^{2}(r)dr^{2}+\frac{r^{2}}{4}\widetilde{g}%
^{2}(r)\{d\theta ^{2}+\sin ^{2}\theta d\phi ^{2}\}+\frac{r^{2}}{4}(d\psi
+\cos \theta d\phi )^{2}  \label{EHIs}
\end{equation}%
where%
\begin{equation}
\begin{array}{c}
\widetilde{f}(r)=\frac{1}{2}(1+\frac{1}{\sqrt{1-\frac{a^{4}}{r^{4}}}}) \\ 
\widetilde{g}(r)=\sqrt{\frac{1}{2}(1+\sqrt{1-\frac{a^{4}}{r^{4}}})}%
\end{array}
\label{EHIsmff}
\end{equation}%
By increasing the parameter $k$ from $0$ to $1$ we obtain series of M2 brane
solutions that provide a smooth transition from Eguchi-Hanson type I M
branes to corresponding branes based on Eguchi-Hanson type II space.

The second M2-brane solution containing Bianchi type IX in its transverse
directions is 
\begin{equation}
ds_{11}^{2}=H(y,r)^{-2/3}\left( -dt^{2}+dx_{1}^{2}+dx_{2}^{2}\right)
+H(y,r)^{1/3}\left( ds_{TN}^{2}+ds_{Bianchi\text{ }IX}^{2}\right)
\label{ds11m2withTN}
\end{equation}%
where we substitute a Taub-NUT space 
\begin{eqnarray}
ds_{TN}^{2} &=&f(y)\left( dy^{2}+y^{2}(d\alpha ^{2}+\sin ^{2}(\alpha )d\beta
^{2})\right) +\left( \frac{(4\mathbf{n})^{2}}{f(y)}\right) \left( d\sigma +%
\frac{1}{2}\cos (\alpha )d\beta \right) ^{2}  \label{dstn4hashi} \\
f(y) &=&\left( 1+\frac{2\mathbf{n}}{y}\right) .
\end{eqnarray}%
for the other half of the transverse space. In this case, after separation
of variables by the relation (\ref{Hyrsep}), we find exactly the same
differential equation for $R(r)$\ as given in equation (\ref{M2eqR}). The
solution of the differential equation for $Y(y)$\ that has a damped
oscillating behavior at infinity up to a constant is 
\begin{equation}
Y(y)=\frac{(-i)\mathcal{W}_{M}(\frac{-ic\mathbf{n}}{\sqrt{2}},1/2,\sqrt{2}%
icy)}{y}  \label{Y2}
\end{equation}%
where $\mathcal{W}_{M}$ \ is the Whittaker function. The final general
solution will be a superposition of all possible solutions in the form 
\begin{equation}
H_{TN\text{ }\otimes Bianchi\text{ }IX}(y,r)=1-i\frac{Q_{M2}}{y}%
\int_{0}^{\infty }dcc^{4}\mathcal{W}_{M}(\frac{-ic\mathbf{n}}{\sqrt{2}},1/2,%
\sqrt{2}icy)R_{c}(r)  \label{HTNAH}
\end{equation}%
as the metric function of M2-brane solution (\ref{ds11m2withTN}). Note that
the Whittaker function in the integrand is pure imaginary, yielding a
real-valued $H_{TN\text{ }\otimes \text{ }Bianchi\text{ }IX}(y,r)$.

Compactifying over the circle parametrized by $\sigma $\ (and noting that $%
\partial /\partial \sigma $\ is a Killing vector) we find the NSNS fields%
\begin{equation}
\begin{array}{rcl}
\Phi & = & \frac{3}{4}\ln \left( \frac{H_{TN\text{ }\otimes Bianchi\text{ }%
IX}^{1/3}}{f}\right) \\ 
B_{\mu \nu } & = & 0%
\end{array}
\label{tnXAHNSNS}
\end{equation}%
\ and RR fields

\begin{equation}
\begin{array}{rcl}
C_{\beta } & = & 2\mathbf{n}\cos (\alpha ) \\ 
A_{tx_{1}x_{2}} & = & H_{TN\text{ }\otimes Bianchi\text{ }IX}^{-1}.%
\end{array}
\label{tnXAHRR}
\end{equation}%
The metric in ten dimensions will be given by 
\begin{eqnarray}
ds_{10}^{2} &=&H_{TN\text{ }\otimes Bianchi\text{ }IX}(y,r)^{-1/2}f^{-1/2}%
\left( -dt^{2}+dx_{1}^{2}+dx_{2}^{2}\right) +H_{TN\text{ }\otimes Bianchi%
\text{ }IX}(y,r)^{1/2}f^{-1/2}ds_{Bianchi\text{ }IX}^{2}  \notag \\
&+&H_{TN\text{ }\otimes Bianchi\text{ }IX}(y,r)^{1/2}f^{1/2}\left(
dy^{2}+y^{2}(d\alpha ^{2}+\sin ^{2}(\alpha )d\beta ^{2})\right) .
\label{ds10tnXAH}
\end{eqnarray}%
This represents a D2$\perp $D6(2) system. To calculate how much
supersymmetry is preserved by this solution in eleven dimensions, we use the
Killing spinor equation (\ref{killingspinoreq}). The Killing spinor equation
is the supersymmetry variation of the gravitino field in $D=11$
supergravity. As we mentioned in section (2), we consider bosonic sector of
eleven dimensional supergravity. Hence to make the vanishing of the
gravitino consistet with the presence of supersymmetry, one has to impose
the constraint that the supersymmetry variations acting on such a bosonic
background cannot restore a non vanishing gravitino. This means that the
supersymmetry variation of the gravitino evaluated in the bosonic sector
must vanish which yields equation (\ref{killingspinoreq}). After a
straightforward but lengthy calculation we have checked that for our brane
solution, the supersymmetry variation vanishes with an arbitrary choice of
one quarter of the components of the spinor $\epsilon$. More specifically,
half of the components of the spinor $\epsilon$ is removed due to the
presence of the brane and another half is removed due to the self-duality of
Bianchi type IX space, so the solution preserves 1/4 of \ the supersymmetry.
We also have explicitly checked that the above 10-dimensional metric, with
the given dilaton, one and three forms, is a solution to the 10-dimensional
supergravity equations of motion.

\begin{figure}[tbp]
%\QTP{Dialog Text}
\centering                                                                   
\begin{minipage}[c]{.3\textwidth}
        \centering
        \includegraphics[width=\textwidth]{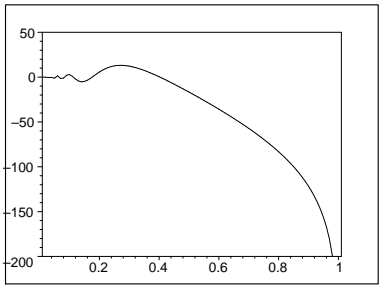}
    \end{minipage}
\caption{{} Numerical solution of equation (\ref{Y0fyforEHAH}) for $%
Y_{k}/10^{5}$ as a function of $\frac{1}{y}$ for non-zero separation
constant $k$. The Eguchi-Hanson parameter $a$ is set to one and so for $%
y\approx a,$ the function $Y_{k}$ diverges and for $y\approx \infty $, it
vanishes.}
\label{fig3}
\end{figure}

\bigskip A third possible M2-brane solution is given by%
\begin{equation}
ds_{11}^{2}=H(y,r)^{-2/3}\left( -dt^{2}+dx_{1}^{2}+dx_{2}^{2}\right)
+H(y,r)^{1/3}\left( ds_{EH}^{2}+ds_{Bianchi\text{ }IX}^{2}\right)
\label{ds11m2withEH}
\end{equation}%
where 
\begin{eqnarray}
ds_{EH}^{2} &=&\frac{y^{2}}{4g(y)}\left[ d\sigma +\cos (\alpha )d\beta %
\right] ^{2}+g(y)dy^{2}+\frac{y^{2}}{4}\left( d\alpha ^{2}+\sin ^{2}(\alpha
)d\beta ^{2}\right)  \label{dseh4} \\
g(y) &=&\left( 1-\frac{a^{4}}{y^{4}}\right) ^{-1}.
\end{eqnarray}%
is the Eguchi-Hanson type II metric.\ In this case, after separation of
variables by the relation (\ref{Hyrsep}), we find the same differential
equation for $R(r)$\ as given in equation (\ref{M2eqR}). The differential
equation for $Y(y)$ is 
\begin{equation}
2y(y^{4}-a^{4})Y_{c}^{\prime \prime }(y)+2(3y^{4}+a^{4})Y_{c}^{\prime
}(y)+c^{2}y^{5}Y_{c}(y)=0.  \label{Y0fyforEHAH}
\end{equation}%
While an analytic closed solution for the differential equation (\ref%
{Y0fyforEHAH}) is not available, the numerical solution shows that it has a
damped oscillating behavior at infinity and that it diverges at $y\simeq a.$%
\ A typical numerical solution of \ $Y_{k}(y)$\ is given in figure \ref{fig3}%
. 
The general solution will be a superposition of all possible solutions in
the form%
\begin{equation}
H_{EH\text{ }\otimes Bianchi\text{ }IX}(y,r)=1+Q_{M2}\int_{0}^{\infty
}dcc^{5}Y_{c}(y)R_{c}(r)  \label{HEHAH0}
\end{equation}%
for the metric function of the M2-brane solution (\ref{ds11m2withEH}), where
we have absorbed a constant into the M2-brane charge.

In this case, by compactification along the $\sigma $ direction of
Eguchi-Hanson metric, we find the NSNS fields%
\begin{equation}
\begin{array}{c}
\Phi =\frac{3}{4}\ln \left\{ \frac{H_{EH\text{ }\otimes Bianchi\text{ }%
IX}^{1/3}w^{2}}{4g}\right\} \\ 
B_{\mu \nu }=0%
\end{array}
\label{NSNSEH}
\end{equation}%
and RR fields%
\begin{equation}
\begin{array}{c}
C_{\beta }=a\cos (\alpha ) \\ 
A_{tx_{1}x_{2}}=\frac{1}{H_{EH\text{ }\otimes Bianchi\text{ }IX}}%
\end{array}
\label{RREH}
\end{equation}%
and metric 
\begin{eqnarray}
ds_{10}^{2} &=&\frac{w}{2}\{H_{EH\text{ }\otimes Bianchi\text{ }%
IX}^{-1/2}g^{-1/2}\left( -dt^{2}+dx_{1}^{2}+dx_{2}^{2}\right) +H_{EH\text{ }%
\otimes Bianchi\text{ }IX}^{1/2}g^{-1/2}ds_{Bianchi\text{ }IX}^{2}  \notag \\
&&+H_{EH\text{ }\otimes Bianchi\text{ }IX}^{1/2}g^{1/2}a^{2}\{dw^{2}+\frac{%
w^{2}}{4g}\left( d\alpha ^{2}+\sin ^{2}\alpha d\beta ^{2}\right) \}\}
\label{ds10AHEH}
\end{eqnarray}%
where $w=\frac{y}{a}.$\ The metric describes an intersecting D2/D6 system
where D2 is localized along the world-volume of the D6-brane and the
world-volume of the D6 brane transverse to D2 is just Bianchi type IX space.
We note that in the large $w$\ limit, the metric (\ref{ds10AHEH}), reduces
to the metric%
\begin{equation}
ds_{10}^{2}=\frac{w}{2}\{-dt^{2}+dx_{1}^{2}+dx_{2}^{2}+ds_{Bianchi\text{ }%
IX}^{2}+a^{2}(dw^{2}+\frac{w^{2}}{4}\left( d\alpha ^{2}+\sin ^{2}\alpha
d\beta ^{2}\right) )\}
\end{equation}%
which is again a 10D locally asymptotically flat metric with Kretchmann
invariant 
\begin{equation}
R_{\mu \nu \rho \sigma }R^{\mu \nu \rho \sigma }=\frac{A(a,r)}{w^{2}}+\frac{%
224}{w^{6}a^{4}}  \label{K2}
\end{equation}%
which vanishes at large $w.$\ $A(a,r)$\ is a complicated function of the
Eguchi-Hanson parameter $a$\ and Bianchi type IX metric function $F(r).$\
All the components of the Riemann tensor in the orthonormal basis approach
zero at $w\rightarrow \infty $.

\bigskip

\bigskip A fourth possible M2-brane solution is given by%
\begin{equation}
ds_{11}^{2}=H(y,r)^{-2/3}\left( -dt^{2}+dx_{1}^{2}+dx_{2}^{2}\right)
+H(y,r)^{1/3}\left( ds_{AH}^{2}+ds_{Bianchi\text{ }IX}^{2}\right)
\label{AHBIX}
\end{equation}%
where%
\begin{equation}
ds_{AH}^{2}=\frac{c^{2}(y)}{y^{2}}dy^{2}+a^{2}(y)\sigma
_{1}^{2}+b^{2}(y)\sigma _{2}^{2}+c^{2}(y)\sigma _{3}^{2}  \label{AHmetric}
\end{equation}%
is the Atiyah-Hitchin metric and $\sigma _{i\text{ }}$are Maurer-Cartan
one-forms on the space with coordinates $\sigma ,\alpha $ and $\beta .$ The
metric functions $a(y),b(y)$ and $c(y)$ can be expressed explicitly in terms
of elliptic integrals $K$ and $E$ \cite{ATM2} and the coordinate $r$ ranges
over the interval $[n\pi ,\infty )$ where the positive number $n$\ is a
constant number with unit of length that is related to NUT charge of metric
at infinity obtained from Atiyah-Hitchin metric. In fact as $y\rightarrow
\infty ,$ the metric (\ref{AHmetric}) reduces to 
\begin{equation}
ds_{AH}^{2}\rightarrow (1-\frac{2n}{y})(dy^{2}+y^{2}d\alpha ^{2}+y^{2}\sin
^{2}\alpha d\beta ^{2})+4n^{2}(1-\frac{2n}{y})^{-1}(d\sigma +\cos \alpha
d\beta )^{2}  \label{reducedAH}
\end{equation}%
which is the well known Euclidean Taub-NUT metric with a negative NUT charge 
$N=-n.$

\begin{figure}[tbp]
\centering                  
\begin{minipage}[c]{.3\textwidth}
        \centering
        \includegraphics[width=\textwidth]{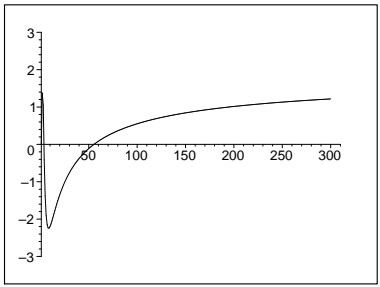}
    \end{minipage}
\caption{ Numerical solution of the radial equation (\ref{R2}) for $\widehat{%
R}_{c}/10^{6}$ as a function of $\frac{1}{r-a_{3}},$ where we set $%
a_{1}=1,a_{2}=2$ and $a_{3}=3$ for simplicity. So for $r\approx a_{3}$ $R$
diverges. Note that the plot is only reliable for large $\frac{1}{r-a_{3}}$. 
}
\label{M2eqRwithposcsq}
\end{figure}

In this case, after separation of variables by the relation (\ref{Hyrsep}),
we find the same differential equation for $R(r)$\ as given in equation (\ref%
{M2eqR}). The differential equation for $Y(y)$ turns out to be 
\begin{equation}
2a(y)b(y)y^{2}Y_{c}^{\prime \prime }(y)+\{2a(y)b(y)y+2a(y)y^{2}\frac{db(y)}{%
dy}+2b(y)y^{2}\frac{da(y)}{dy}\}Y_{c}^{\prime
}(y)+c^{2}a(y)b(y)c^{2}(y)Y_{c}(y)=0.
\end{equation}%
While an analytic closed solution for the differential equation (\ref%
{Y0fyforEHAH}) is not available, we have found numerical solutions of this
equation in paper \cite{ATM2}. The numerical solutions show for $y\simeq
n\pi ,$ function $Y_{c}(y)$ diverges logarithmically by $\ln (\frac{1}{%
y-n\pi })$ and for large $y,$ it has oscillating damped behavior$.$\ A
typical numerical solution of \ $Y_{k}(y)$\ is given in figure \ref{fig3}. \
So, the general solution for M2 metric function will be a superposition of
all possible solutions in the form%
\begin{equation}
H_{AH\text{ }\otimes Bianchi\text{ }IX}(y,r)=1+Q_{M2}\int_{0}^{\infty
}dcc^{5}R_{c}(r)Y_{c}(y)
\end{equation}%
Finally, the fifth possible M2-brane solution is given by%
\begin{equation}
ds_{11}^{2}=H(r_{1},r_{2})^{-2/3}\left( -dt^{2}+dx_{1}^{2}+dx_{2}^{2}\right)
+H(r_{1},r_{2})^{1/3}\left( ds_{Bianchi\text{ }IX_{1}}^{2}+ds_{Bianchi\text{ 
}IX_{2}}^{2}\right)   \label{AHAH}
\end{equation}%
where $ds_{Bianchi\text{ }IX_{1}}^{2}$\ and $ds_{Bianchi\text{ }IX_{2}}^{2}$%
\ are given by two copies of (\ref{BIX}) with coordinate systems $%
(r_{1},\theta _{1},\phi _{1},\psi _{1})$\ and $(r_{2},\theta _{2},\phi
_{2},\psi _{2})$ and two sets of metric functions $(A_{1},A_{2},A_{3})$ and $%
(\widehat{A}_{1},\widehat{A}_{2},\widehat{A}_{3})$, respectively. \ In this
case, after separation of variables by the relation 
\begin{equation}
H(r_{1},r_{2})=1+Q_{M2}R_{c}(r_{1})\widehat{R}_{c}(r_{2})  \label{sepofvar}
\end{equation}%
\ we find two differential equations%
\begin{equation}
2r_{1}A_{1}A_{2}A_{3}\frac{d^{2}R_{c}(r_{1})}{dr_{1}^{2}}%
+\{6A_{1}A_{2}A_{3}+r_{1}(A_{2}A_{3}\frac{dA_{1}}{dr_{1}}+A_{3}A_{1}\frac{%
dA_{2}}{dr_{1}}+A_{1}A_{2}\frac{dA_{3}}{dr_{1}})\}\frac{dR_{c}(r_{1})}{dr_{1}%
}-c^{2}r_{1}\sqrt{A_{1}A_{2}A_{3}}R_{c}(r_{1})=0  \label{R1}
\end{equation}%
\ 
\begin{equation}
2r_{2}\widehat{A}_{1}\widehat{A}_{2}\widehat{A}_{3}\frac{d^{2}\widehat{R}%
_{c}(r_{2})}{dr_{2}^{2}}+\{6\widehat{A}_{1}\widehat{A}_{2}\widehat{A}%
_{3}+r_{2}(\widehat{A}_{2}\widehat{A}_{3}\frac{d\widehat{A}_{1}}{dr_{2}}+%
\widehat{A}_{3}\widehat{A}_{1}\frac{d\widehat{A}_{2}}{dr_{2}}+\widehat{A}_{1}%
\widehat{A}_{2}\frac{d\widehat{A}_{3}}{dr_{2}})\}\frac{d\widehat{R}%
_{c}(r_{2})}{dr_{2}}+c^{2}r_{2}\sqrt{\widehat{A}_{1}\widehat{A}_{2}\widehat{A%
}_{3}}\widehat{R}_{c}(r_{2})=0.  \label{R2}
\end{equation}%
The differential equation for $R_{c}(r_{1})$\ is the same as equation (\ref%
{M2eqR}) and typical solutions are presented in figures \ref{M2R123} - \ref%
{M2R121000}. As before, an analytic closed-form solution for the first
differential equation (\ref{R2}) is not available. However the numerical
solution presented in figure \ref{M2eqRwithposcsq}\ shows that it has
diverging behavior at $r\simeq a_{3}$\ and damped oscillating behavior at
infinity.

The general solution will be a superposition of all possible solutions in
the form%
\begin{equation}
H_{Bianchi\text{ }IX_{1}\text{ }\otimes \text{ }Bianchi\text{ }%
IX_{2}}(r_{1},r_{2})=1+Q_{M2}\int_{0}^{\infty }dcc^{5}R_{c}(r_{1})\widehat{R}%
_{c}(r_{2})  \label{Hah2}
\end{equation}%
where we have absorbed additional constants into the M2-brane charge.

To summarize, all M2-brane solutions with a Bianchi type IX space in the
transverse geometry preserve 1/4 of the supersymmetry. In all of these
M2-brane solutions half of the supersymmetry is removed due to the presence
of the M2-brane and another half is removed due to the self-duality of
Bianchi type IX space.

\section{A Second set of M2-brane solutions}

\label{sec:sc}{\Large \ \ \ \ \ }

A different set of M2-brane solutions can be obtained by reversing the sign
of the separation constant $c^{2}$ in the separated differential equations
for $Y(y)$ and $R(r).$ As an example, by taking $c\rightarrow i\widetilde{c}$
in the separable equations of \ embedded Bianchi type IX case, we find the
solution of the differential equation for $\widetilde{Y}(y)$ as 
\begin{equation}
\widetilde{Y}(y)=\frac{K_{1}(\frac{\widetilde{c}}{\sqrt{2}}y)}{y}
\end{equation}%
where $K_{1}$\ is the modified Bessel function, diverging at $y=0$ and
vanishing at infinity. The differential equation for $\widetilde{R}(r)$ is
given by

\begin{equation}
2rA_{1}A_{2}A_{3}\frac{d^{2}\widetilde{R}_{\widetilde{c}}(r)}{dr^{2}}%
+\{6A_{1}A_{2}A_{3}+r(A_{2}A_{3}\frac{dA_{1}}{dr}+A_{3}A_{1}\frac{dA_{2}}{dr}%
+A_{1}A_{2}\frac{dA_{3}}{dr})\}\frac{d\widetilde{R}_{\widetilde{c}}(r)}{dr}+%
\widetilde{c}^{2}r\sqrt{A_{1}A_{2}A_{3}}\widetilde{R}_{\widetilde{c}}(r)=0.
\label{M2eqRsc}
\end{equation}%
Although the above equation does not have any analytic closed solution, we
can solve it numerically. Typical solutions are presented in figure \ref%
{M2eqRwithposcsq}. The final general solution is the second M2-brane
solution in (\ref{ds11m2}), and is a superposition of all possible solutions 
\begin{equation}
\widetilde{H}_{Bianchi\text{ }IX}(y,r)=1+Q_{M2}\int_{0}^{\infty }d\widetilde{%
c}\widetilde{c}^{4}\frac{K_{1}(\frac{\widetilde{c}}{\sqrt{2}}y)}{y}%
\widetilde{R}_{\widetilde{c}}(r)  \label{scBIX}
\end{equation}%
absorbing a possible constant into the charge\textbf{\ }$Q_{M2}$.

The other two alternative solutions for the Taub-NUT $\otimes $ Bianchi type
IX and Eguchi-Hanson $\otimes $ Bianchi type IX could be derived easily
similar to the above case and so we do not present them here. In the Bianchi
type IX $\otimes $\ Bianchi type IX case with metric function (\ref{Hah2}),
the transformation $c\rightarrow i\widetilde{c}$\ in (\ref{R1}) and (\ref{R2}%
), merely interchanges $R_{c}(r_{1})\rightarrow \widehat{R}_{\widetilde{c}%
}(r_{1})$\ and $\widehat{R}_{c}(r_{2})\rightarrow R_{\widetilde{c}}(r_{2})$
and so yields no new solution.

\section{Embedding Bianchi type IX space in an M5-brane metric}

\label{sec:M5}

The eleven dimensional M5-brane metric with an embedded Bianchi type IX
metric has the following form%
\begin{equation}
\begin{array}{c}
ds_{11}^{2}=H(y,r)^{-1/3}\left(
-dt^{2}+dx_{1}^{2}+dx_{2}^{2}+dx_{3}^{2}+dx_{4}^{2}+dx_{5}^{2}\right)  \\ 
+H(y,r)^{2/3}\left( dy^{2}+ds_{Bianchi\text{ }IX}^{2}\right) 
\end{array}
\label{ds11m5p}
\end{equation}%
with field strength components%
\begin{equation}
\begin{array}{c}
F_{\psi \theta \phi y}=\frac{1}{16}\alpha r^{3}\sin \theta \sqrt{%
A_{1}(r)A_{2}(r)A_{3}(r)}\frac{\partial H}{\partial r} \\ 
F_{r\theta \phi \psi }=\frac{1}{16}\alpha r^{3}\sin \theta \frac{\partial H}{%
\partial y}.%
\end{array}
\label{FScompo}
\end{equation}%
We consider the M5-brane which corresponds to $\alpha =+1$;\ the $\alpha =-1$%
\ case corresponds to an anti-M5 brane.

The metric (\ref{ds11m5p}) is a solution to the eleven dimensional
supergravity equations provided $H\left( y,r\right) $ is a solution to the
differential equation 
\begin{equation}
2A_{1}A_{2}A_{3}\frac{\partial ^{2}H}{\partial r^{2}}+\{A_{2}A_{3}\frac{%
dA_{1}}{dr}+A_{1}A_{2}\frac{dA_{3}}{dr}+A_{1}A_{3}\frac{dA_{2}}{dr}+\frac{6}{%
r}A_{1}A_{2}A_{3}\}\frac{\partial H}{\partial r}+2\sqrt{A_{1}A_{2}A_{3}}%
\frac{\partial ^{2}H}{\partial y^{2}}=0.  \label{HeqforM5}
\end{equation}%
This equation is straightforwardly separable. Substituting 
\begin{equation}
H(y,r)=1+Q_{M5}Y(y)R(r)
\end{equation}%
where $Q_{M5}$ is the charge on the M5-brane. The solution of the
differential equation for $Y(y)$ is 
\begin{equation}
Y(y)=\cos (\frac{c}{\sqrt{2}}y+\varsigma )
\end{equation}%
and the differential equation for $R(r)$ is given by the equation (\ref%
{M2eqR}). The numerical solution of this equation is presented in figures %
\ref{M2R123},\ref{M2R12500} and \ref{M2R121000} for different values of $%
a_{1},a_{2}$ and $a_{3}$. The final solution is a superposition of all
possible solutions in the form of%
\begin{equation}
H(y,r)=1+Q_{M5}\int_{0}^{\infty }dcc^{2}\cos (\frac{c}{\sqrt{2}}y)R_{c}(r)
\label{M5}
\end{equation}%
as the metric function of M5-brane solution (\ref{ds11m5p}), where we absorb
the constant into the M5-brane charge.

A different M5-brane solution can be obtained by reversing the sign of the
separation constant $c^{2}$\ in the separated differential equations
obtained from (\ref{HeqforM5}).\ In this case, by taking $c\rightarrow i%
\widetilde{c}$\ , we find another solution in the form of%
\begin{equation}
\widetilde{H}(y,r)=1+Q_{M5}\int_{0}^{\infty }d\widetilde{c}\widetilde{c}%
^{2}e^{-\widetilde{c}y}\widetilde{R}_{\widetilde{c}}(r)  \label{M5ss}
\end{equation}%
where numerical plot of the function $\widetilde{R}_{\widetilde{c}}$\ is
given in figure \ref{M2eqRwithposcsq}. \ Although this is formally a
solution, the integral in (\ref{M5ss}) is not convergent for all values of $%
y $. To make the integral convergent for $y<0$, one can replace $e^{-%
\widetilde{c}y}$\ by $e^{-\widetilde{c}\left| y\right| }$, but only at the
price of introducing a source term at $y=0$\ in the corresponding Laplace
equation for $\ \widetilde{H}(y,r).$

\section{Decoupling limits}

\bigskip \label{sec:declim}

At low energies, the dynamics of the D2 brane decouple from the bulk, with
the region close to the D6 brane corresponding to a range of energy scales
governed by the IR fixed point \cite{DecouplingLim}. For D2 branes localized
on D6 branes, this corresponds in the field theory to a vanishing mass for
the fundamental hyper-multiplets. Near the D2 brane horizon ($H\gg 1$), the
field theory limit is given by 
\begin{equation}
g_{YM2}^{2}=g_{s}\ell _{s}^{-1}=\text{fixed.}  \label{gymFTlimit}
\end{equation}%
In this limit the gauge couplings in the bulk go to zero, so the dynamics
there decouple. In each of our cases above, the radial coordinates are also
scaled such that 
\begin{equation}
Y=\frac{y}{\ell _{s}^{2}}~~,~~~~U=\frac{r}{\ell _{s}^{2}}
\label{YUdecoupling1}
\end{equation}%
are fixed. As an example we note that this will change the harmonic function
of the D6 brane in the TN $\otimes $ Bianchi IX case to the following
(recall that the asymptotic radius of the 11th dimension is $R_{\infty }=4%
\mathbf{n}=g_{s}\ell _{s}$) 
\begin{equation}
f(y)=\left( 1+\frac{2\mathbf{n}}{y}\right) =\left( 1+\frac{g_{s}\ell _{s}}{2y%
}\right) =\left( 1+\frac{g_{s}}{2\ell _{s}Y}\right) =\left( 1+\frac{%
g_{YM2}^{2}N_{6}}{2Y}\right) =f(Y)  \label{F2}
\end{equation}%
where we generalize to the case of $N_{6}$ D6 branes, giving the factor of $%
N_{6}$ in the final line above. A similar analysis shows for the EH metric,
we get 
\begin{equation}
g(y)\rightarrow \left( 1-\frac{A^{4}}{Y^{4}}\right) =g(Y)  \label{EHfU}
\end{equation}%
where $a$ has been rescaled to $a=A\ell _{s}^{-2}$.

All of the D2 harmonic functions from the above solutions can be shown to
scale as $H(Y,U)=\ell _{s}^{-4}h(Y,U)$. This form causes the D2-brane to
warp the ALE region and the asymptotically flat region of the D6-brane
geometry. The $h(Y,U)$'s are easily calculated; as an example, the TN $%
\otimes $ Bianchi IX function is given by\bigskip 
\begin{equation}
h_{TN\text{ }\otimes Bianchi\text{ }IX}(Y,U)=i\frac{32\pi ^{2}N_{2}g_{YM}^{2}%
}{Y}\int_{0}^{\infty }dPP^{4}\mathcal{W}_{M}(\frac{-i(\frac{Pg_{YM2}^{2}}{4})%
}{\sqrt{2}},1/2,\sqrt{2}iPY)R_{P}(U)  \label{hTNBIX}
\end{equation}%
\bigskip\ where \bigskip we used $\ell _{p}=g_{s}^{1/3}\ell _{s}$ to rewrite 
\begin{equation}
Q_{M2}=32\pi ^{2}N_{2}\ell _{p}^{6}=32\pi ^{2}N_{2}g_{YM2}^{2}\ell _{s}^{8}.
\label{Qm2value}
\end{equation}%
In equation (\ref{hTNBIX}), $R_{P}(U)$ is the solution of 
\begin{eqnarray}
2UA_{1}A_{2}A_{3}\frac{d^{2}R_{P}(U)}{dU^{2}} &+&%
\{6A_{1}A_{2}A_{3}+U(A_{2}A_{3}\frac{dA_{1}}{dU}+A_{3}A_{1}\frac{dA_{2}}{dU}%
+A_{1}A_{2}\frac{dA_{3}}{dU})\}\frac{dR_{P}(U)}{dU}  \notag \\
&-&P^{2}U\sqrt{A_{1}A_{2}A_{3}}R_{P}(U)=0  \label{ww}
\end{eqnarray}%
where $A_{i}(U)=1-\frac{\gamma _{i}^{4}}{U^{4}}$ and we rescaled the
integration variable $c$ and Bianchi IX parameters $a_{i}$ by $P/\ell
_{s}^{2}$ and $\ell _{s}^{2}\gamma _{i}$, respectively. Even when full
analytic forms of $H(y,r)$ are not available, we can show that $H(Y,U)=\ell
_{s}^{-4}h(Y,U)$ in the decoupling limit, due to the general forms of $H(y,r)
$ we obtained above. As an example, we can see the ten-dimensional metric (%
\ref{ds10tnXAH}) scales as\bigskip 
\begin{eqnarray}
\frac{ds_{10}^{2}}{\ell _{s}^{2}} &=&h_{TN\text{ }\otimes Bianchi\text{ }%
IX}(Y,U)^{-1/2}f(U)^{-1/2}\left( -dt^{2}+dx_{1}^{2}+dx_{2}^{2}\right) + 
\notag \\
&+&h_{TN\text{ }\otimes Bianchi\text{ }IX}(Y,U)^{1/2}f(U)^{-1/2}\{\frac{%
dU^{2}}{\sqrt{F(U)}}+\frac{U^{2}}{4}\sqrt{F(U)}\left( \frac{\sigma _{1}^{2}}{%
1-\frac{\gamma _{1}^{4}}{U^{4}}}+\frac{\sigma _{2}^{2}}{1-\frac{\gamma
_{2}^{4}}{U^{4}}}+\frac{\sigma _{3}^{2}}{1-\frac{\gamma _{3}^{4}}{U^{4}}}%
\right) \}  \notag \\
&+&h_{TN\text{ }\otimes Bianchi\text{ }IX}(y,r)^{1/2}f^{1/2}\left(
dY^{2}+Y^{2}(d\alpha ^{2}+\sin ^{2}(\alpha )d\beta ^{2})\right) .
\end{eqnarray}%
and there is only an overall normalization factor of $\ell _{s}^{2}$ in the
above metric which is the expected result for a solution that is a
supergravity dual of a quantum field theory.

\bigskip

\bigskip We now want to perform an analysis of the decoupling limits of the
solutions presented above. \ At low energies, the dynamics of IIA NS5-branes
will decouple from the bulk. Near the NS5-brane horizon ($H>>1$), we are
interested in the behavior of the NS5-branes in the limit where string
coupling vanishes 
\begin{equation}
g_{s}\rightarrow 0  \label{gym1}
\end{equation}%
and 
\begin{equation}
\ell _{s}=\text{ fixed.}  \label{gym2}
\end{equation}%
\ In these limits, we rescale the radial coordinates such that they can be
kept fixed 
\begin{equation}
Y=\frac{y}{g_{s}\ell _{s}^{2}}~,~U=\frac{r}{g_{s}\ell _{s}^{2}}.
\label{yrrescale}
\end{equation}

We can show the harmonic functions (\ref{M5}) and (\ref{M5ss}) for the
NS5-branes to rescale according to $H(Y,U)=g_{s}^{-2}h(Y,U).$ For example,
the harmonic function (\ref{M5}) becomes 
\begin{eqnarray}
H(Y,U) &=&\pi N_{5}g_{s}\int_{0}^{\infty }\frac{1}{g_{s}}dP(\frac{P^{2}}{%
g_{s}^{2}})\cos \left( \frac{PY}{\sqrt{2}}\right) R_{P}(U)  \notag \\
&=&\frac{h(Y,U)}{g_{s}^{2}}
\end{eqnarray}%
where we have rescaled $c=Pg_{s}^{-1}\ell _{s}^{-2}$ so that $h(Y,U)$ has no 
$g_{s}$ dependence and we have used $\ell _{p}=g_{s}^{1/3}\ell _{s}$ to
rewrite $Q_{M5}=\pi N_{5}\ell _{p}^{3}=\pi N_{5}g_{s}\ell _{s}^{3}.$

In the limit of vanishing $g_{s}$\ with fixed $l_{s}$\ (as we did in (\ref%
{gym1}) and (\ref{gym2})), the decoupled free theory on NS5-branes should be
a little string theory \cite{shiraz} (i.e. a 6-dimensional non-gravitational
theory in which modes on the 5-brane interact amongst themselves, decoupled
from the bulk).

\section{Conclusion}

By embedding Bianchi type IX space into M-theory, we have found new classes
of \ 2-brane and 5-brane solutions to $D=11$\ supergravity. These exact
solutions are new M2- and M5-brane metrics with metric functions (\ref%
{gensolutionnearreqPi}), (\ref{HTNAH}), (\ref{HEHAH0}), (\ref{Hah2}), (\ref%
{scBIX}), (\ref{M5}) and (\ref{M5ss}) -- these are the main results of this
paper. The important point is that all previously found M2 and M5 solutions
based on embedded Taub-NUT, Eguchi Hanson type II and Atiyah-Hitchin spaces
are special cases of our new solutions based on triaxial Biachi type IX
space. Moreover we found series of M2 (and simiarly for M5) brane solutions
by increasing the parameter $k$ from $0$ to $1$ in (\ref{special}) that
provide a smooth transition from Eguchi-Hanson type I M branes to
corresponding branes based on Eguchi-Hanson type II space. The common
feature of both solutions is that the brane function is a convolution of an
exponentially decaying `radial' function (for both branes) with a damped
oscillating one. The `radial' function vanishes far from the branes and
diverges logarithmically near the brane core. \ The same logarithmic
divergence near the brane happens in embedding of \ Eguchi-Hanson II metric
in M-theory where the divergence is milder than $\frac{1}{r},$\ as in the
case of embedding Taub-NUT space. Indeed, all of these properties of our
solutions are similar to those previously obtained \cite{CGMM2,ATM2} for the
embedding of Taub-NUT, Eguchi-Hanson II and Atiyah-Hitchin spaces.

Our solutions preserve 1/4 of the supersymmetry due to the self-dual
character of the Bianchi type IX metric. This is in contrast to earlier
brane solutions of this type \cite{CGMM2}, for which M.supersymmetry could
only be preserved for NUT-like transverse metrics; their bolt counterparts
did not preserve any supersymmetry. Dimensional reduction of the M2
solutions to ten dimensions gives us intersecting IIA D2/D6 \ configurations
that preserve 1/4 of the supersymmetry.

Finally we considered the decoupling limit of our solutions. In the case of
embedded TN$_{4}$ in M2 solution \cite{hashi}, when the D2 brane decouples
from the bulk, the theory on the brane is 3 dimensional $\mathcal{N}=4$ $SU($%
N$_{2})$ super Yang-Mills (with eight supersymmetries) coupled to N$_{6}$\
massless hypermultiplets \cite{pelc}. This point is obtained from dual field
theory and since some of our solutions preserve the same amount of
supersymmetry, a similar dual field description should be attainable.
\bigskip

{\Large Acknowledgments}

I would like to thank Professor R. Mann for helpful discussions and valuable
comments.

\end{document}